\documentclass[final,english]{bullsrsl}
\usepackage{setspace}

\usepackage[latin1]{inputenc}
\usepackage{graphicx}
\usepackage{sidecap}
\usepackage{tikz}

\usepackage[T1]{fontenc}

\usepackage{natbib}          

\begin{document}
\title{The fastest hot subdwarfs revisited}

%
%
%
%

\author[affil={1}, corresponding]{Ulrich}{Heber}
\author[affil={2}]{MUCHFUSS collaboration}{}

% ...

%
\affiliation[1]{Dr. Karl Remeis-Sternwarte \& ECAP, 
Astronomical Institute, FAU, University of Erlangen-Nuremberg, Germany}
\affiliation[2]{MUCHFUSS collaboration: see \citet{2015A&A...577A..26G}}

\correspondance{ulrich.heber@fau.de}

\maketitle

\begin{abstract}
Hyper-velocity stars (HVS) are enigmatic objects because they are travelling so fast that they escape from the Galaxy. Among hot subdwarfs, only one such star is known, the He-sdO US\,708.
The Hyper-MUCHFUSS collaboration provided additional HVS candidates. Here we revisit the fastest candidates including US\,708 by analysing optical spectra and spectral energy distributions using a new grid of tailored model atmospheres and report preliminary results.
The sample is dominated by H-rich subdwarfs and their distribution in the Kiel diagram appears to be bimodal for the sdB stars but otherwise fits canonical evolutionary models well. \textit{Gaia}'s proper motion measurements allowed a precise kinematic investigation to be made. It turns out that all previously proposed HVS candidates are actually bound to the Galaxy, except US\,708. The original candidate sample turns out to belong to an extreme halo population. The scarcity of available observations of US\,708 calls for space-based UV and IR photometry as well as high precision radial velocity measurements. 
\end{abstract}

\section{The search for sdO stars in the SDSS and the discovery of the hyper-velocity subdwarf US\,708}
 
\begin{figure}
\centering
\includegraphics[width=0.95\textwidth]{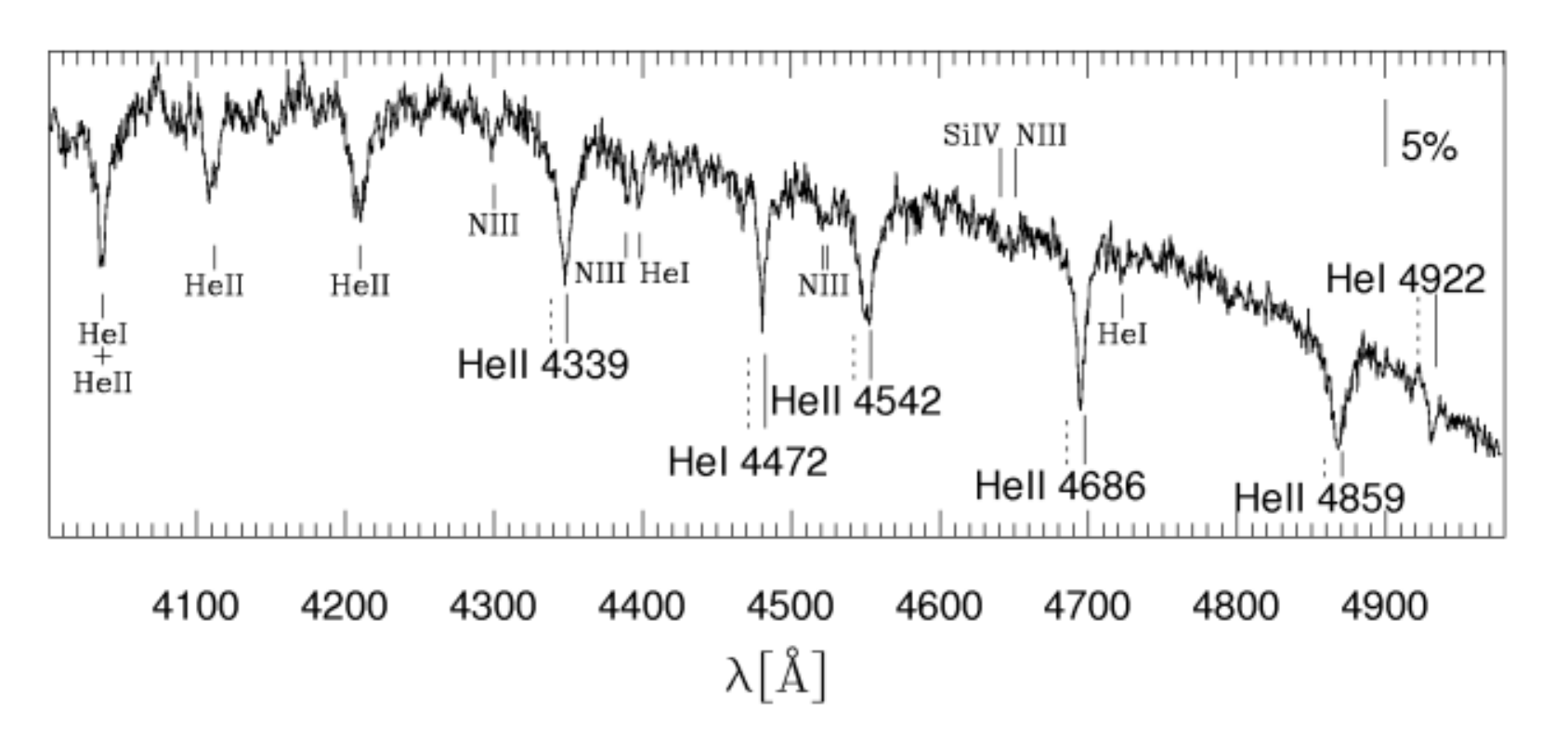}

\caption{KECK LRIS spectrum of the He-sdO 
US\,708.  Note the large redshift of the spectral lines \citep{Hirsch2006}. Reproduced with permission by H. Hirsch.}\label{fig:keck_lris}
\end{figure}

\begin{SCfigure}
\includegraphics[width=0.8\textwidth]{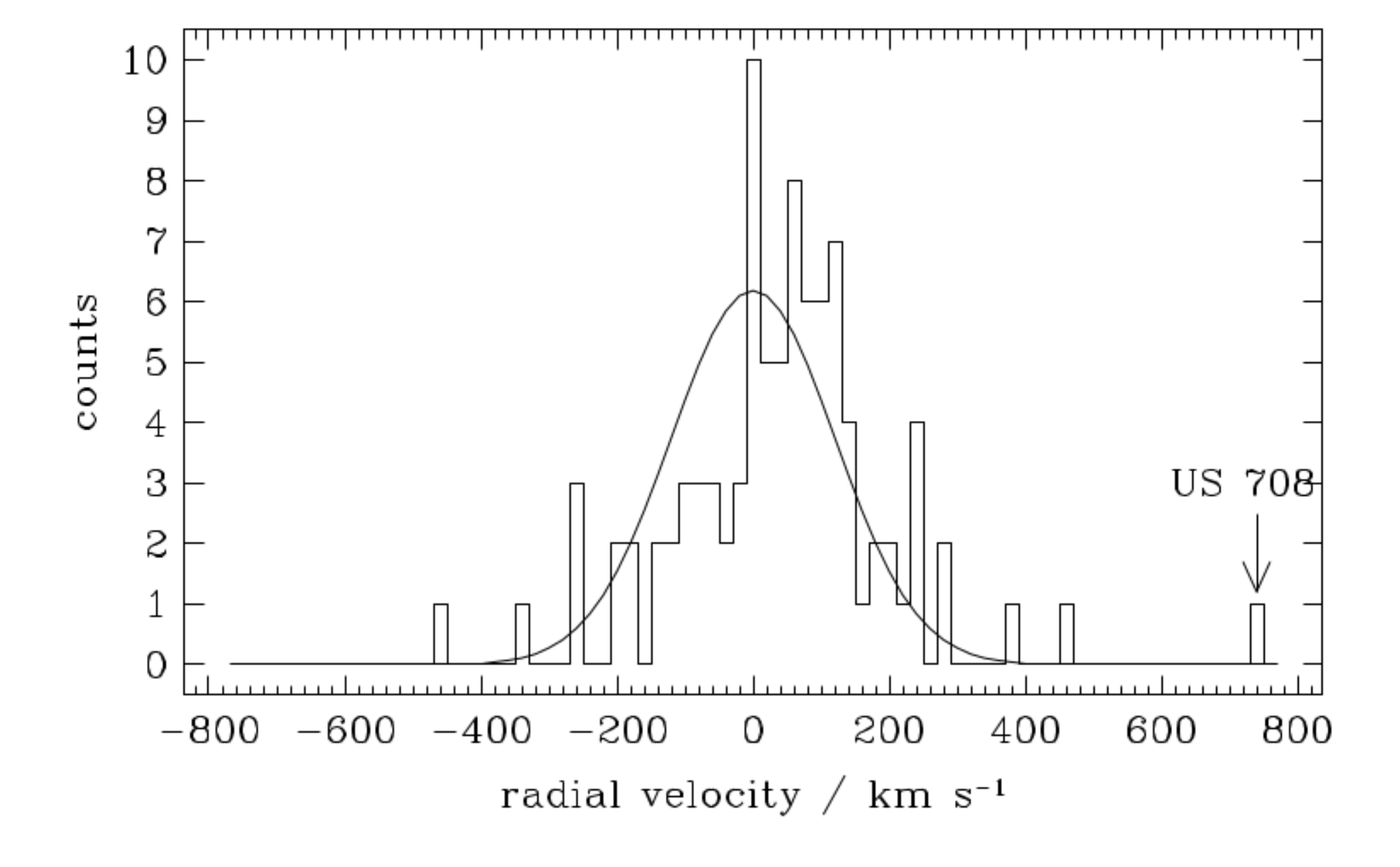}
\caption{ Distribution of radial velocities of a sample of sdO stars from the SDSS \citep{Hirsch2006}. Reproduced with permission by H. Hirsch.}\label{fig:keck_rv}
\end{SCfigure}

Three hyper-velocity stars were discovered in 2005\citep{2005ApJ...622L..33B,2005ApJ...634L.181E} the second one (HVS\,2) turned out to be a He-sdO star \citep[][see Fig. \ref{fig:keck_lris}]{2005A&A...444L..61H}. This serendipitous discovery came from
the diploma project of Heiko Hirsch in Bamberg \citep{Hirsch2006} who was given the task to search for sdO stars in the SDSS spectral data release 4. 
By visual inspection of 13\,000 colour-selected targets he discovered 118 sdO stars, $\approx$50\% turned out to be helium-rich (He-sdOs). 
SDSSJ093320.86+441705.4 stood out from the rest by its very high radial velocity (see Fig. \ref{fig:keck_rv}). No contribution of hydrogen Balmer lines to the He {\sc II} Pickering series is detectable. The strength of the nitrogen lines indicates that US\,708 belongs to the N-subclass of He-sdO stars\citep{2007A&A...462..269S}. 
well above the local Galactic escape velocity. Follow-up spectroscopy at the Keck telescope confirmed the radial velocity measurement \citep{2005A&A...444L..61H}, proving that the star is unbound to the Galaxy. The star was first mentioned as a faint blue object in a deep photographic survey of Kapteyn's field SA\,29 by \citet{1982ApJS...48...51U} and, henceforth, has been referred to as US\,708, accordingly. The search for hot subdwarfs stars was then intensified \citep{2007AN....328..657H} and the HYPER-MUCHFUSS project was launched to identify the fastest ones \citep{2011A&A...527A.137T}. 

Further studies focused on the place of origin and the ejection mechanism. 
When hyper-velocity stars were first discovered \citep{2005ApJ...622L..33B}, ejection by tidal disruption of a binary by the supermasssive black hole in the Galactic centre was readily suggested as ejection mechanism \citep{1988Natur.331..687H}. 
Proper motion measurements are required to calculate the tangential velocity. However, ground-based data were mostly too inaccurate to be useful. Nevertheless, for US\,708
\citet{2015Sci...347.1126G} derived sufficiently precise proper motions using Pan-STARRS data to trace back the trajectory of US\,708 to the Galactic plane. They could exclude the Galactic centre at the 5 sigma level of significance, but suggest an origin in the Galactic plane. This excludes the Hills mechanism and called for an alternative ejection scenario. \citet{2015Sci...347.1126G} suggested that US\,708 is the surviving companion in a  double detonation binary supernova scenario. Helium mass transfer from the He-sdO to a C/O white dwarf companion began when the binary orbital period decreased to P$<$20 min.
Helium accumulates at the surface of the white dwarf until a thermonuclear explosion of the helium layer sets in which then triggers the explosion of the C/O core, creating a SN Ia event.
The surviving companion (US\,708) was ejected travelling beyond the escape velocity from the Galaxy. The loss of a significant amount of envelope mass had reduced the mass of US\,708 to just 0.3M$_\odot$.
\citet{2020A&A...641A..52N} revisited the binary supernova model making use of \textit{Gaia} data. His investigations suggested that the most likely current mass of US\,708  lies between
0.34 M$_\odot$ and 0.37 M$_\odot$ in a Chandrasekhar-mass SN event, significantly higher than the mass of 0.3 M$_\odot$ assumed by \citet{2015Sci...347.1126G}.
An observational determination of the mass for US\,708 is still pending.

In the course of the Hyper-MUCHFUSS project, candidates were selected based on unusually high RVs and proper motions. Because the latter taken from various catalogs can be notoriously incorrect, proper motions were derived individually from photographic plates \citep[e.g.][]{2011A&A...527A.137T,2017A&A...601A..58Z}. This led to the discovery of a dozen candidate HVS subdwarfs including the He-sdO SDSSJ205030.39-061957.8 (J2050-063) spectroscopically similar to US\,708 \citep{2017A&A...601A..58Z}. \citet{2016ApJ...821L..13N} discovered a binary sdB star and argued that it may originate from a tidally destroyed satellite galaxy.

\section{\textit{Gaia} -- the game changer}

\textit{Gaia}'s second data release became a game changer, because precise proper motions became available for all of the several hundreds of HVS candidates of various spectral types suggested in the literature.
The new results were sobering, because almost all candidates selected from high tangential velocities were eliminated by the \textit{Gaia} astrometry \citep{2018MNRAS.479.2789B}. Essentially, we are left with the about 4 dozen HVS discovered from their exceedingly high RVs.

\subsection{US\,708 in the \textit{Gaia} era} \label{sect:us708}

\textit{Gaia} DR2 provided proper motions of superior quality to the ones used by \citet{2015Sci...347.1126G}. To calculate the tangential motion the distance is required as well.  Since \textit{Gaia} can not provide the parallax of US\,708 to a sufficient precision, we studied its spectral energy distribution (SED) to derive the angular diameter and the reddening based on the atmospheric parameters from \citep{2015Sci...347.1126G}. To derive the spectrophotmetric distance we adopt the low mass predicted by the binary supernova scenario. 

\begin{SCfigure}
\includegraphics[angle=0,width=0.6\textwidth]{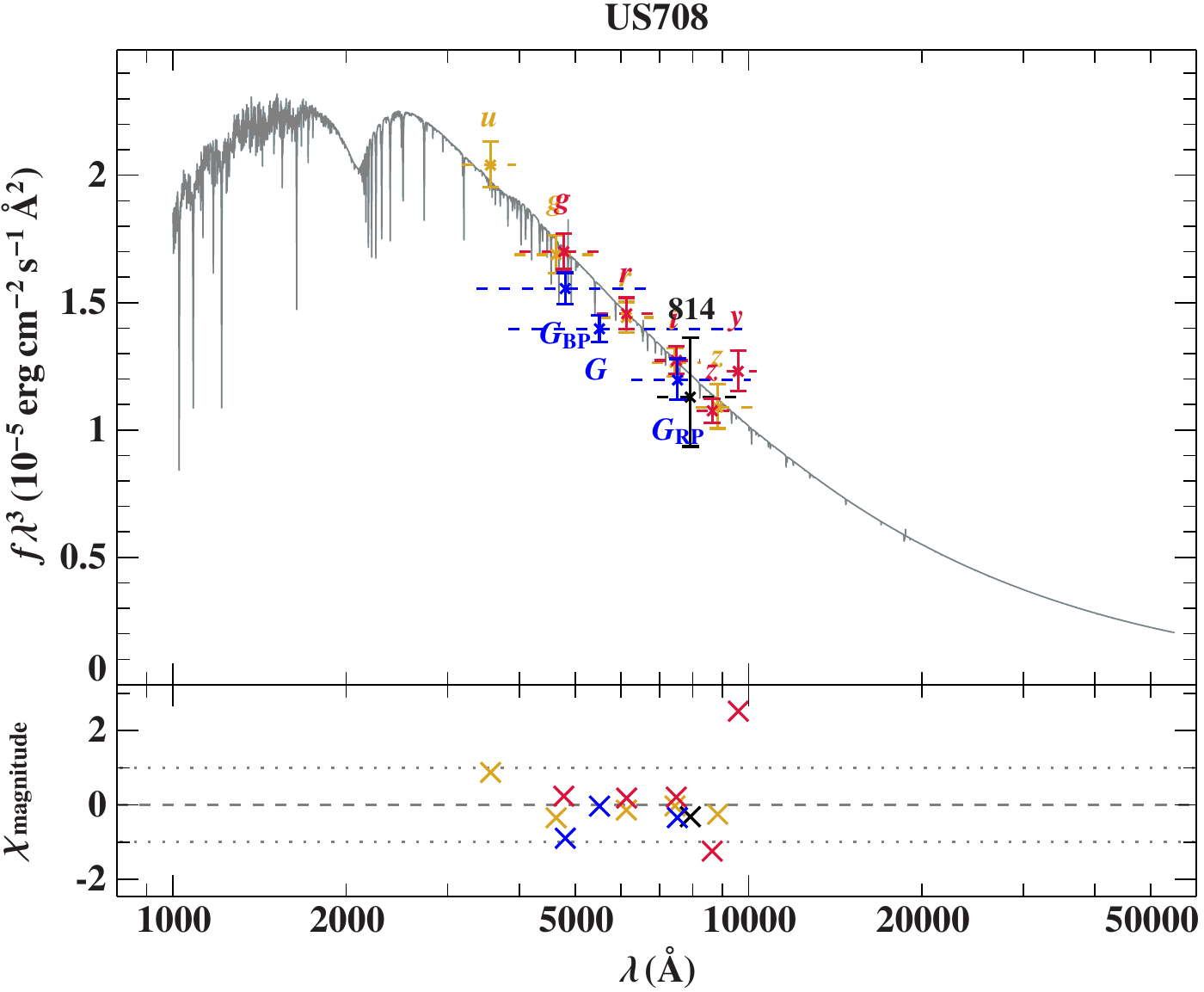}
\caption{US\,708: comparison of synthetic (solid gray) and observed photometry
(blue: \textit{Gaia} EDR3; yellow Pan-STARRS DR1, red: SDSS DR12). The top
panel shows the SED. The colored data points are filter-averaged fluxes
that were converted from observed magnitudes; the solid gray line represents the best fitting model. The flux is multiplied by the wavelength to the power
of three. The bottom panel
shows the residuals, $\chi$, i.e. the difference between synthetic and
observed magnitudes divided by the corresponding uncertainties.}\label{fig:us708_sed}
\end{SCfigure}

We constructed the SED (see Fig. \ref{fig:us708_sed}) by collecting photometry from various data bases and fitted them with synthetic SEDs \citep[for details see ][]{2018OAst...27...35H} to derive the angular diameter of the star of $\log(\Theta\,\mathrm{(rad)})$ = $-12.152 \pm 0.009$  as well as the interstellar reddening by $E(B-V)$ = $0.041 \pm 0.015$\,mag. Combining these results with the surface gravity of $\log\,g = 5.69 \pm 0.09$ derived by \citet{2015Sci...347.1126G} places US\,708 at 8.2 $\pm$ 1 kpc, if its mass is 0.3 M$_\odot$ or 10.3 $\pm$ 1.2 kpc if its mass were canonical (0.47 M$_\odot$).

The place of origin in the Galactic disk is determined by integrating the equation of motion in a Galactic potential \citep[for details see][]{2013A&A...549A.137I} back to the last disk crossing. 
Today's best velocity estimates for US\,708 based on \textit{Gaia}'s DR2  are a Galactic restframe velocity of  v$_{\rm grf }$=994\,km\,s$^{-1}$, and an ejection velocity from the Galactic disk of v$_{\rm ej, disk}$=897 km\,s$^{-1}$ \citep{2020A&A...641A..52N}. These values are somewhat lower than those found by \citet{2015Sci...347.1126G} which relaxes the tension on the binary supernova ejection model, which remains a viable scenario. 

\subsection{The fastest hot subdwarfs revisited}

In an ongoing project we also revisit the fastest hot subdwarf stars from the MUCHFUSS project \citep{2015A&A...577A..26G} starting with new spectroscopic analyses, and adding spectrophotometric as well as kinematic analyses. 

\begin{figure}
\centering
\vspace*{1cm}
\includegraphics[width=0.9\textwidth]{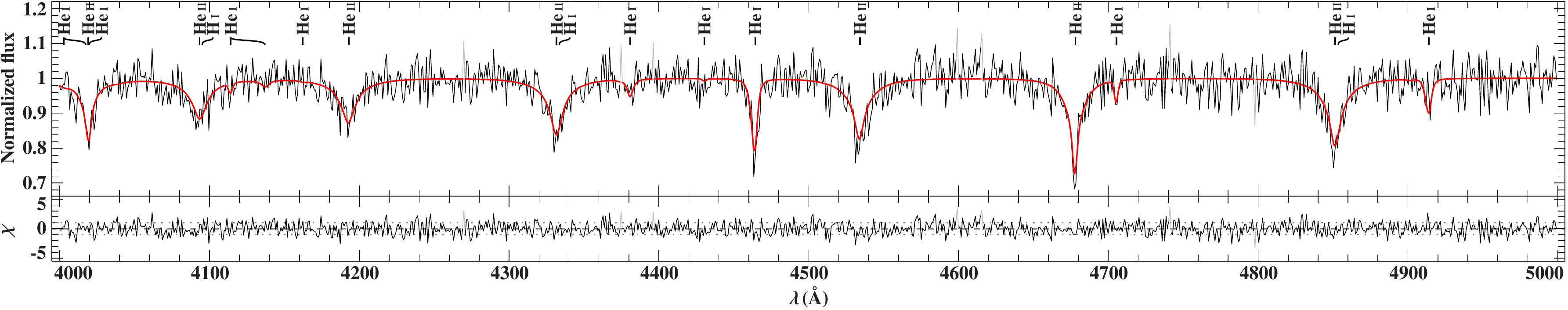}
\caption{Fit of a SDSS spectrum of J2050-063. 
The red line is the best fitting synthetic spectrum. The lower panel gives the residual $\chi$.}\label{fig:j2050}

\end{figure}

\subsubsection{Quantitative spectral analysis: New model grid \& analysis strategy}

A new grid of atmospheric models has become available that replaces the LTE grid of \citet{2000A&A...363..198H}. It uses a hybrid LTE/NLTE approach \citep[for details see][]{2011JPhCS.328a2015P,2021A&A...650A.102I}. The temperature--density stratification is calculated using Kurucz' \textit{ATLAS12} code for an average subdwarf chemical composition.  \textit{DETAIL} and \textit{SURFACE} were used to calculate NLTE departure coefficients for H and He and synthesise the spectrum. The global fitting procedure of \citet{2014A&A...565A..63I} replaces the selective ones previously used.  

\begin{SCfigure}
\includegraphics[width=0.73\textwidth]{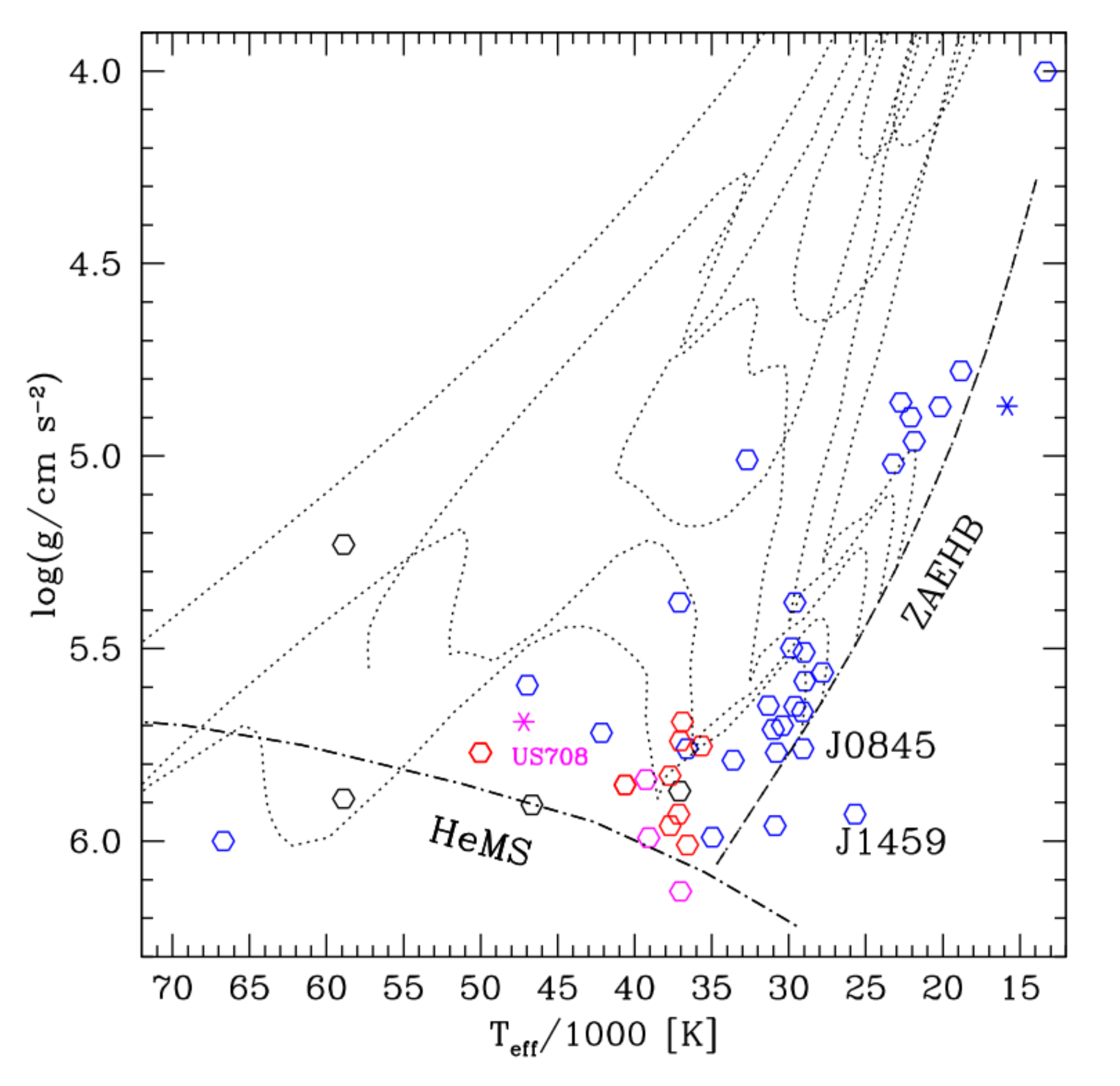}
\caption{Position of the fastest hot subdwarfs in the Kiel (T$_\textrm{eff}$--log\,g) diagram. H-rich objects are marked in blue, eHe-sdO and iHe-sdO stars are shown in magenta and red, respectively. The four sdOs of solar He content are marked as black symbols, two of which are very hot ($\approx$60kK) and evolved.
The zero age EHB and evolutionary tracks are from \citet{1993ApJ...419..596D} for [Fe/H]=-2.29. The dashed-dotted line marks the helium main sequence.}\label{fig:kiel}
\end{SCfigure}

We use these models and the analysis strategy to re-analyse the spectra of the fastest hot subdwarfs from the MUCHFUSS project. As an example we show the fit of a single SDSS spectrum of the US708 twin SDSSJ205030.39-061957.8 in Fig. \ref{fig:j2050}.
 The sample is dominated by hydrogen-rich sdB stars plus a dozen helium-rich sdO stars. Interestingly, nine of them belong to the intermediate He-sdO subclass. The results are shown in the Kiel (T$_\textrm{eff}$--log\,g) diagram (Fig. \ref{fig:kiel}). As can be seen the sdB stars form two groups; a relatively cool group at T$_\textrm{eff}$=20\ldots 25 kK and a hotter one at T$_\textrm{eff}$=28\ldots 35 kK.

\subsubsection{Kinematics}

\begin{SCfigure}
\includegraphics[width=0.7\textwidth]{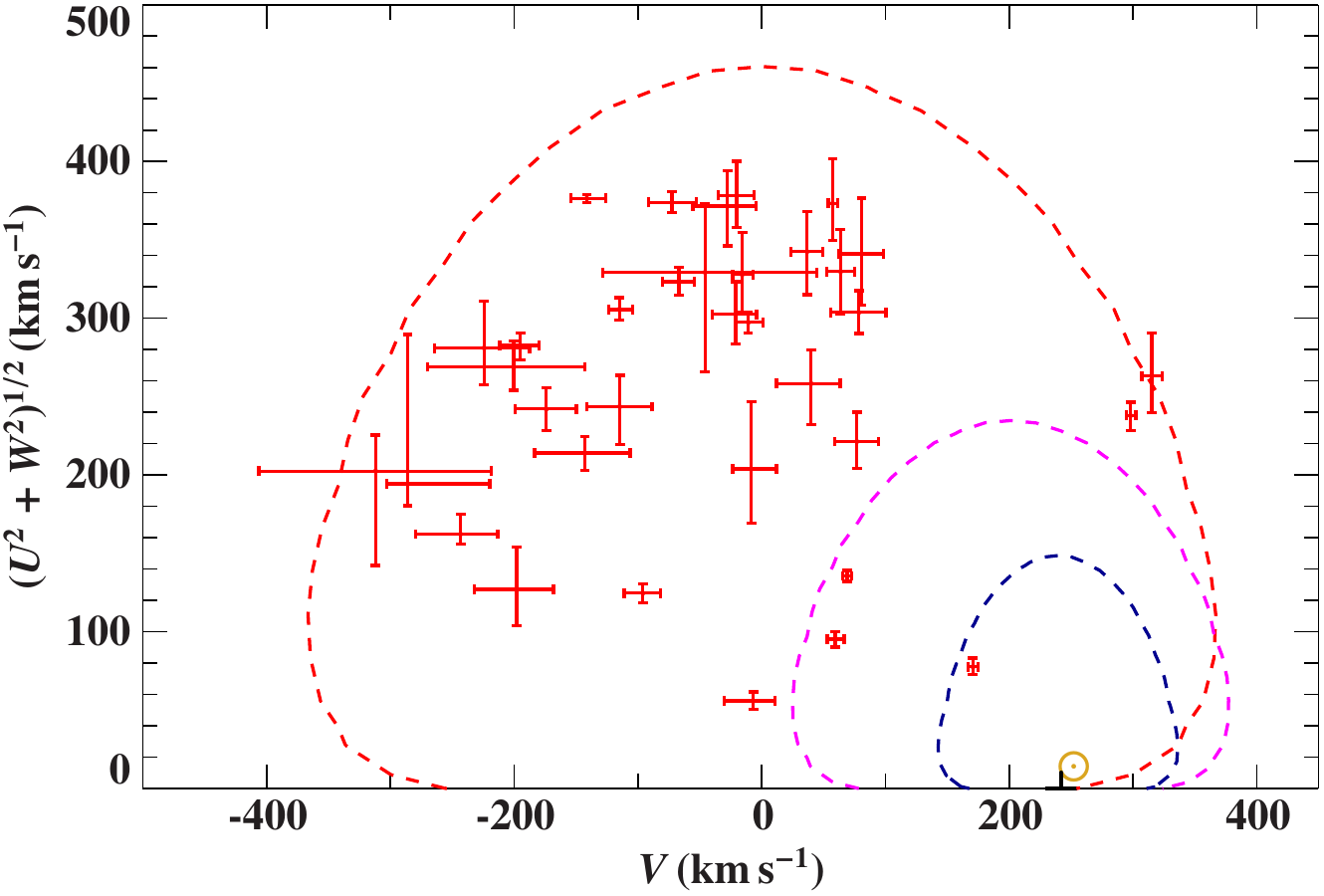}
\caption{The Toomre diagram for the sample stars (red crosses with 1 $\sigma$ error bars):  the velocity component V is measured in the direction of Galactic rotation, U towards the Galactic centre, and W perpendicular
to the Galactic plane. The Sun is marked by a yellow $\odot$. Contours (3$\sigma$, from \citet{2003A&A...409..523R}) mark the thin disk (blue), thick disk (magenta), and halo population (red).  
}\label{fig:toomre}
\end{SCfigure}

High precision proper motions were provided by the early third data release of \textit{Gaia} \citep{2021A&A...649A...2L}.
As for US\,708 the stars are too far away for Gaia parallaxes to be useful. Therefore, spectroscopic distances were calculated in a similar way as for US\,708 (see Sect. \ref{sect:us708} and Fig. \ref{fig:us708_sed}). 
The SEDs were constructed and analysed to derive the angular diameters and interstellar extinctions. The distances are then calculated from the spectroscopically determined surface gravities adopting the canonical mass of 0.47 M$_\odot$. Radial velocities from the spectral fits were added to derive the space velocity vector. 
The resulting Galactic restframe velocities are lower than the local Galactic escape velocities indicating that no star of the sample is unbound to the Galaxy. It is, therefore, likely, that the stars belong to an old halo population. In order to test this conclusion we compare their space velocity to predictions from Milky Way models \citep{2003A&A...409..523R} in the Toomre diagram (Fig. \ref{fig:toomre}). Indeed, the distribution of stars is consistent with halo membership. US\,708 remains the only known HVS star among hot subdwarfs. 

\section{Conclusion}

We present preliminary results from an ongoing extensive spectroscopic, spectrophotometric, and kinematic  analysis of the fastest hot subdwarfs of the MUCHFUSS project including the 
hyper-velocity subdwarf US\,708. \textit{Gaia}'s proper motion measurements confirmed that US\,708 can not originate from the Galactic centre and, therefore, Hills' slingshot mechanism can be excluded. The revised space velocities of US\,708  are somewhat smaller than previously estimated relaxing the tension on the binary supernova ejection model for US\,708 \citep{2015Sci...347.1126G,2020A&A...641A..52N}. 

The sample is dominated by H-rich subdwarfs and its distribution in the Kiel diagram appears to be bimodal for the sdB stars but otherwise fits canonical evolutionary models well. Except US\,708 all stars turn out to be bound to the Galaxy. Their distribution in the kinematic Toomre diagram reveals that it is a sample of halo stars. Despite of its uniqueness, observations of US\,708 are scarce. Photometry is available in the optical regime only. Ultraviolet as well as infrared photometry is still lacking to characterise the star. This calls for HST observations for the UV and JWST for the IR. No high precision radial velocity measurements are available to test RV variability. 






\begin{acknowledgments}
The author thanks the German Science foundation (DFG) or funding the MUCHUSS project through grants HE1356/45-1\&2.
\end{acknowledgments}
\bibliographystyle{bullsrsl-en}

\bibliography{heber_sdob10.bib}

\begin{thebibliography}{24}
\providecommand{\natexlab}[1]{#1}
\providecommand{\url}[1]{#1}
\providecommand{\urlprefix}{URL }

\bibitem[{{Boubert} et~al.(2018){Boubert}, {Guillochon}, {Hawkins}, {Ginsburg},
  {Evans} and {Strader}}]{2018MNRAS.479.2789B}
{Boubert}, D., {Guillochon}, J., {Hawkins}, K., {Ginsburg}, I., {Evans}, N.~W.
  and {Strader}, J. (2018) {Revisiting hypervelocity stars after Gaia DR2}.
\newblock MNRAS, 479(2), 2789--2795.
\newblock \url{https://doi.org/10.1093/mnras/sty1601}.

\bibitem[{{Brown} et~al.(2005){Brown}, {Geller}, {Kenyon} and
  {Kurtz}}]{2005ApJ...622L..33B}
{Brown}, W.~R., {Geller}, M.~J., {Kenyon}, S.~J. and {Kurtz}, M.~J. (2005)
  {Discovery of an Unbound Hypervelocity Star in the Milky Way Halo}.
\newblock ApJL, 622(1), L33--L36.
\newblock \url{https://doi.org/10.1086/429378}.

\bibitem[{{Dorman} et~al.(1993){Dorman}, {Rood} and
  {O'Connell}}]{1993ApJ...419..596D}
{Dorman}, B., {Rood}, R.~T. and {O'Connell}, R.~W. (1993) {Ultraviolet
  Radiation from Evolved Stellar Populations. I. Models}.
\newblock ApJ, 419, 596.
\newblock \url{https://doi.org/10.1086/173511}.

\bibitem[{{Edelmann} et~al.(2005){Edelmann}, {Napiwotzki}, {Heber},
  {Christlieb} and {Reimers}}]{2005ApJ...634L.181E}
{Edelmann}, H., {Napiwotzki}, R., {Heber}, U., {Christlieb}, N. and {Reimers},
  D. (2005) {HE 0437-5439: An Unbound Hypervelocity Main-Sequence B-Type Star}.
\newblock ApJL, 634(2), L181--L184.
\newblock \url{https://doi.org/10.1086/498940}.

\bibitem[{{Geier} et~al.(2015{\natexlab{a}}){Geier}, {F{\"u}rst}, {Ziegerer},
  {Kupfer}, {Heber}, {Irrgang}, {Wang}, {Liu}, {Han}, {Sesar}, {Levitan},
  {Kotak}, {Magnier}, {Smith}, {Burgett}, {Chambers}, {Flewelling}, {Kaiser},
  {Wainscoat} and {Waters}}]{2015Sci...347.1126G}
{Geier}, S., {F{\"u}rst}, F., {Ziegerer}, E., {Kupfer}, T., {Heber}, U.,
  {Irrgang}, A., {Wang}, B., {Liu}, Z., {Han}, Z., {Sesar}, B., {Levitan}, D.,
  {Kotak}, R., {Magnier}, E., {Smith}, K., {Burgett}, W.~S., {Chambers}, K.,
  {Flewelling}, H., {Kaiser}, N., {Wainscoat}, R. and {Waters}, C.
  (2015{\natexlab{a}}) {The fastest unbound star in our Galaxy ejected by a
  thermonuclear supernova}.
\newblock Science, 347(6226), 1126--1128.
\newblock \url{https://doi.org/10.1126/science.1259063}.

\bibitem[{{Geier} et~al.(2015{\natexlab{b}}){Geier}, {Kupfer}, {Heber},
  {Schaffenroth}, {Barlow}, {{\O}stensen}, {O'Toole}, {Ziegerer}, {Heuser},
  {Maxted}, {G{\"a}nsicke}, {Marsh}, {Napiwotzki}, {Br{\"u}nner}, {Schindewolf}
  and {Niederhofer}}]{2015A&A...577A..26G}
{Geier}, S., {Kupfer}, T., {Heber}, U., {Schaffenroth}, V., {Barlow}, B.~N.,
  {{\O}stensen}, R.~H., {O'Toole}, S.~J., {Ziegerer}, E., {Heuser}, C.,
  {Maxted}, P.~F.~L., {G{\"a}nsicke}, B.~T., {Marsh}, T.~R., {Napiwotzki}, R.,
  {Br{\"u}nner}, P., {Schindewolf}, M. and {Niederhofer}, F.
  (2015{\natexlab{b}}) {The catalogue of radial velocity variable hot
  subluminous stars from the MUCHFUSS project}.
\newblock A\&A, 577, A26.
\newblock \url{https://doi.org/10.1051/0004-6361/201525666}.

\bibitem[{{Heber} et~al.(2018){Heber}, {Irrgang} and
  {Schaffenroth}}]{2018OAst...27...35H}
{Heber}, U., {Irrgang}, A. and {Schaffenroth}, J. (2018) {Spectral energy
  distributions and colours of hot subluminous stars}.
\newblock Open Astronomy, 27(1), 35--43.
\newblock \url{https://doi.org/10.1515/astro-2018-0008}.

\bibitem[{{Heber} et~al.(2000){Heber}, {Reid} and
  {Werner}}]{2000A&A...363..198H}
{Heber}, U., {Reid}, I.~N. and {Werner}, K. (2000) {Spectral analysis of multi
  mode pulsating sdB stars. II. Feige 48, KPD 2109+4401 and PG 1219+534}.
\newblock A\&A, 363, 198--207.

\bibitem[{{Hills}(1988)}]{1988Natur.331..687H}
{Hills}, J.~G. (1988) {Hyper-velocity and tidal stars from binaries disrupted
  by a massive Galactic black hole}.
\newblock Nature, 331(6158), 687--689.
\newblock \url{https://doi.org/10.1038/331687a0}.

\bibitem[{{Hirsch}(2006)}]{Hirsch2006}
{Hirsch}, H. (2006) {Hei\ss e unterleuchtkr\"aftige Sterne aus dem Sloan
  Digital Sky Survey}.
\newblock {Diploma thesis}, {Friedrich--Alexander Universit\"at},
  {Erlangen--N\"urnberg}.

\bibitem[{{Hirsch} et~al.(2007){Hirsch}, {Heber} and
  {O'Toole}}]{2007AN....328..657H}
{Hirsch}, H., {Heber}, U. and {O'Toole}, S. (2007) {Hot subluminous O stars
  from the SDSS}.
\newblock Astronomische Nachrichten, 328(7), 657.

\bibitem[{{Hirsch} et~al.(2005){Hirsch}, {Heber}, {O'Toole} and
  {Bresolin}}]{2005A&A...444L..61H}
{Hirsch}, H.~A., {Heber}, U., {O'Toole}, S.~J. and {Bresolin}, F. (2005) {US
  708 - an unbound hyper-velocity subluminous O star}.
\newblock A\&A, 444(3), L61--L64.
\newblock \url{https://doi.org/10.1051/0004-6361:200500212}.

\bibitem[{{Irrgang} et~al.(2021){Irrgang}, {Geier}, {Heber}, {Kupfer},
  {El-Badry} and {Bloemen}}]{2021A&A...650A.102I}
{Irrgang}, A., {Geier}, S., {Heber}, U., {Kupfer}, T., {El-Badry}, K. and
  {Bloemen}, S. (2021) {A proto-helium white dwarf stripped by a substellar
  companion via common-envelope ejection. Uncovering the true nature of a
  candidate hypervelocity B-type star}.
\newblock A\&A, 650, A102.
\newblock \url{https://doi.org/10.1051/0004-6361/202038757}.

\bibitem[{{Irrgang} et~al.(2014){Irrgang}, {Przybilla}, {Heber}, {B{\"o}ck},
  {Hanke}, {Nieva} and {Butler}}]{2014A&A...565A..63I}
{Irrgang}, A., {Przybilla}, N., {Heber}, U., {B{\"o}ck}, M., {Hanke}, M.,
  {Nieva}, M.~F. and {Butler}, K. (2014) {A new method for an objective,
  {\ensuremath{\chi}}$^{2}$-based spectroscopic analysis of early-type stars.
  First results from its application to single and binary B- and late O-type
  stars}.
\newblock A\&A, 565, A63.
\newblock \url{https://doi.org/10.1051/0004-6361/201323167}.

\bibitem[{{Irrgang} et~al.(2013){Irrgang}, {Wilcox}, {Tucker} and
  {Schiefelbein}}]{2013A&A...549A.137I}
{Irrgang}, A., {Wilcox}, B., {Tucker}, E. and {Schiefelbein}, L. (2013) {Milky
  Way mass models for orbit calculations}.
\newblock A\&A, 549, A137.
\newblock \url{https://doi.org/10.1051/0004-6361/201220540}.

\bibitem[{{Lindegren} et~al.(2021){Lindegren}, {Klioner}, {Hern{\'a}ndez},
  {Bombrun}, {Ramos-Lerate}, {Steidelm{\"u}ller}, {Bastian}, {Biermann}, {de
  Torres}, {Gerlach}, {Geyer}, {Hilger}, {Hobbs}, {Lammers}, {McMillan},
  {Stephenson}, {Casta{\~n}eda}, {Davidson}, {Fabricius}, {Gracia-Abril},
  {Portell}, {Rowell}, {Teyssier}, {Torra}, {Bartolom{\'e}}, {Clotet},
  {Garralda}, {Gonz{\'a}lez-Vidal}, {Torra}, {Abbas}, {Altmann}, {Anglada
  Varela}, {Balaguer-N{\'u}{\~n}ez}, {Balog}, {Barache}, {Becciani}, {Bernet},
  {Bertone}, {Bianchi}, {Bouquillon}, {Brown}, {Bucciarelli}, {Busonero},
  {Butkevich}, {Buzzi}, {Cancelliere}, {Carlucci}, {Charlot}, {Cioni},
  {Crosta}, {Crowley}, {del Peloso}, {del Pozo}, {Drimmel}, {Esquej}, {Fienga},
  {Fraile}, {Gai}, {Garcia-Reinaldos}, {Guerra}, {Hambly}, {Hauser},
  {Jan{\ss}en}, {Jordan}, {Kostrzewa-Rutkowska}, {Lattanzi}, {Liao}, {Licata},
  {Lister}, {L{\"o}ffler}, {Marchant}, {Masip}, {Mignard}, {Mints}, {Molina},
  {Mora}, {Morbidelli}, {Murphy}, {Pagani}, {Panuzzo}, {Pe{\~n}alosa Esteller},
  {Poggio}, {Re Fiorentin}, {Riva}, {Sagrist{\`a} Sell{\'e}s}, {Sanchez
  Gimenez}, {Sarasso}, {Sciacca}, {Siddiqui}, {Smart}, {Souami}, {Spagna},
  {Steele}, {Taris}, {Utrilla}, {van Reeven} and
  {Vecchiato}}]{2021A&A...649A...2L}
{Lindegren}, L., {Klioner}, S.~A., {Hern{\'a}ndez}, J., {Bombrun}, A.,
  {Ramos-Lerate}, M., {Steidelm{\"u}ller}, H., {Bastian}, U., {Biermann}, M.,
  {de Torres}, A., {Gerlach}, E., {Geyer}, R., {Hilger}, T., {Hobbs}, D.,
  {Lammers}, U., {McMillan}, P.~J., {Stephenson}, C.~A., {Casta{\~n}eda}, J.,
  {Davidson}, M., {Fabricius}, C., {Gracia-Abril}, G., {Portell}, J., {Rowell},
  N., {Teyssier}, D., {Torra}, F., {Bartolom{\'e}}, S., {Clotet}, M.,
  {Garralda}, N., {Gonz{\'a}lez-Vidal}, J.~J., {Torra}, J., {Abbas}, U.,
  {Altmann}, M., {Anglada Varela}, E., {Balaguer-N{\'u}{\~n}ez}, L., {Balog},
  Z., {Barache}, C., {Becciani}, U., {Bernet}, M., {Bertone}, S., {Bianchi},
  L., {Bouquillon}, S., {Brown}, A.~G.~A., {Bucciarelli}, B., {Busonero}, D.,
  {Butkevich}, A.~G., {Buzzi}, R., {Cancelliere}, R., {Carlucci}, T.,
  {Charlot}, P., {Cioni}, M. R.~L., {Crosta}, M., {Crowley}, C., {del Peloso},
  E.~F., {del Pozo}, E., {Drimmel}, R., {Esquej}, P., {Fienga}, A., {Fraile},
  E., {Gai}, M., {Garcia-Reinaldos}, M., {Guerra}, R., {Hambly}, N.~C.,
  {Hauser}, M., {Jan{\ss}en}, K., {Jordan}, S., {Kostrzewa-Rutkowska}, Z.,
  {Lattanzi}, M.~G., {Liao}, S., {Licata}, E., {Lister}, T.~A., {L{\"o}ffler},
  W., {Marchant}, J.~M., {Masip}, A., {Mignard}, F., {Mints}, A., {Molina}, D.,
  {Mora}, A., {Morbidelli}, R., {Murphy}, C.~P., {Pagani}, C., {Panuzzo}, P.,
  {Pe{\~n}alosa Esteller}, X., {Poggio}, E., {Re Fiorentin}, P., {Riva}, A.,
  {Sagrist{\`a} Sell{\'e}s}, A., {Sanchez Gimenez}, V., {Sarasso}, M.,
  {Sciacca}, E., {Siddiqui}, H.~I., {Smart}, R.~L., {Souami}, D., {Spagna}, A.,
  {Steele}, I.~A., {Taris}, F., {Utrilla}, E., {van Reeven}, W. and
  {Vecchiato}, A. (2021) {Gaia Early Data Release 3. The astrometric solution}.
\newblock A\&A, 649, A2.
\newblock \url{https://doi.org/10.1051/0004-6361/202039709}.

\bibitem[{{N{\'e}meth} et~al.(2016){N{\'e}meth}, {Ziegerer}, {Irrgang},
  {Geier}, {F{\"u}rst}, {Kupfer} and {Heber}}]{2016ApJ...821L..13N}
{N{\'e}meth}, P., {Ziegerer}, E., {Irrgang}, A., {Geier}, S., {F{\"u}rst}, F.,
  {Kupfer}, T. and {Heber}, U. (2016) {An Extremely Fast Halo Hot Subdwarf Star
  in a Wide Binary System}.
\newblock ApJL, 821(1), L13.
\newblock \url{https://doi.org/10.3847/2041-8205/821/1/L13}.

\bibitem[{{Neunteufel}(2020)}]{2020A&A...641A..52N}
{Neunteufel}, P. (2020) {Exploring velocity limits in the thermonuclear
  supernova ejection scenario for hypervelocity stars and the origin of US
  708}.
\newblock A\&A, 641, A52.
\newblock \url{https://doi.org/10.1051/0004-6361/202037792}.

\bibitem[{{Przybilla} et~al.(2011){Przybilla}, {Nieva} and
  {Butler}}]{2011JPhCS.328a2015P}
{Przybilla}, N., {Nieva}, M.-F. and {Butler}, K. (2011) {Testing common
  classical LTE and NLTE model atmosphere and line-formation codes for
  quantitative spectroscopy of early-type stars}.
\newblock In Journal of Physics Conference Series, vol. 328 of \emph{Journal of
  Physics Conference Series}, p. 012015.
\newblock \url{https://doi.org/10.1088/1742-6596/328/1/012015}.

\bibitem[{{Robin} et~al.(2003){Robin}, {Reyl{\'e}}, {Derri{\`e}re} and
  {Picaud}}]{2003A&A...409..523R}
{Robin}, A.~C., {Reyl{\'e}}, C., {Derri{\`e}re}, S. and {Picaud}, S. (2003) {A
  synthetic view on structure and evolution of the Milky Way}.
\newblock A\&A, 409, 523--540.
\newblock \url{https://doi.org/10.1051/0004-6361:20031117}.

\bibitem[{{Stroeer} et~al.(2007){Stroeer}, {Heber}, {Lisker}, {Napiwotzki},
  {Dreizler}, {Christlieb} and {Reimers}}]{2007A&A...462..269S}
{Stroeer}, A., {Heber}, U., {Lisker}, T., {Napiwotzki}, R., {Dreizler}, S.,
  {Christlieb}, N. and {Reimers}, D. (2007) {Hot subdwarfs from the ESO
  supernova Ia progenitor survey. II. Atmospheric parameters of subdwarf O
  stars}.
\newblock A\&A, 462(1), 269--280.
\newblock \url{https://doi.org/10.1051/0004-6361:20065564}.

\bibitem[{{Tillich} et~al.(2011){Tillich}, {Heber}, {Geier}, {Hirsch},
  {Maxted}, {G{\"a}nsicke}, {Marsh}, {Napiwotzki}, {{\O}stensen} and
  {Scholz}}]{2011A&A...527A.137T}
{Tillich}, A., {Heber}, U., {Geier}, S., {Hirsch}, H., {Maxted}, P.~F.~L.,
  {G{\"a}nsicke}, B.~T., {Marsh}, T.~R., {Napiwotzki}, R., {{\O}stensen}, R.~H.
  and {Scholz}, R.~D. (2011) {The Hyper-MUCHFUSS project: probing the Galactic
  halo with sdB stars}.
\newblock A\&A, 527, A137.
\newblock \url{https://doi.org/10.1051/0004-6361/201015539}.

\bibitem[{{Usher} et~al.(1982){Usher}, {Mattson} and
  {Warnock}}]{1982ApJS...48...51U}
{Usher}, P.~D., {Mattson}, D. and {Warnock}, I., A. (1982) {Faint blue objects
  at high galactic latitude. II. Palomar Schmidt field centered on SA 29.}
\newblock ApJS, 48, 51--71.
\newblock \url{https://doi.org/10.1086/190767}.

\bibitem[{{Ziegerer} et~al.(2017){Ziegerer}, {Heber}, {Geier}, {Irrgang},
  {Kupfer}, {F{\"u}rst} and {Schaffenroth}}]{2017A&A...601A..58Z}
{Ziegerer}, E., {Heber}, U., {Geier}, S., {Irrgang}, A., {Kupfer}, T.,
  {F{\"u}rst}, F. and {Schaffenroth}, J. (2017) {Spectroscopic twin to the
  hypervelocity sdO star US 708 and three fast sdB stars from the
  Hyper-MUCHFUSS project}.
\newblock A\&A, 601, A58.
\newblock \url{https://doi.org/10.1051/0004-6361/201730437}.

\end{thebibliography}
\end{document}